\documentclass{Interspeech2024}

\usepackage{color}
\usepackage{cite}
\usepackage{subcaption}
\usepackage{multirow}

\interspeechcameraready

\title{Song Data Cleansing for End-to-End Neural Singer Diarization Using Neural Analysis and Synthesis Framework}
\name{Hokuto Munakata, Ryo Terashima, Yusuke Fujita}

\address{LY Corporation, Japan}
\email{\{hokuto.munakata, ryo.terashima, yusuke.fujita\}@lycorp.co.jp}


\begin{document}

\maketitle

\begin{abstract}
We propose a data cleansing method that utilizes a neural analysis and synthesis (NANSY++) framework to train an end-to-end neural diarization model (EEND) for singer diarization.
Our proposed model converts song data with choral singing which is commonly contained in popular music and unsuitable for generating a simulated dataset to the solo singing data.
This cleansing is based on NANSY++, which is a framework trained to reconstruct an input non-overlapped audio signal.
We exploit the pre-trained NANSY++ to convert choral singing into clean, non-overlapped audio.
This cleansing process mitigates the mislabeling of choral singing to solo singing and helps the effective training of EEND models even when the majority of available song data contains choral singing sections.
We experimentally evaluated the EEND model trained with a dataset using our proposed method using annotated popular duet songs.
As a result, our proposed method improved $14.8$ points in diarization error rate.

\end{abstract}

\section{Introduction}
Singer information analysis is an important aspect of music information processing~\cite{goto2014singing, humphrey2018introduction, su2018vocal}.
Duet or group songs, where multiple singers take turns or sing simultaneously, are common in popular music.
Our purpose is singer diarization, a task to predict ``who sings when'' in the multiple-singer songs.
This task can be applied to displaying the lyrics of each singer separately on karaoke systems.

The formulation of the singer diarization task is similar to the well-known speaker diarization task, which has been investigated for a long time~\cite{bredin2020pyannote, landini2022bayesian, garcia2017speaker, medennikov2020target}.
In speaker diarization, end-to-end neural diarization (EEND) is appealing for its performance and the ability to handle overlapped speech directly~\cite{fujita2019end, horiguchi2020end, kinoshita2021integrating}.
The performance of the EEND model is supported by pre-training with a large dataset.
The training data for this model consists of pairs of speech data and their corresponding speaker labels, i.e., speaker identities and utterance duration.
For the training, simulated dataset generation methods~\cite{fujita2019end, landini2022simulated} have been commonly used because they can easily generate a large dataset without annotation cost.
In this method, clean isolated speech data of two or more speakers are mixed and the speaker labels are obtained from the individual isolated speech.
These simulated datasets are supported by a large public dataset of conversational speech data~\cite{callhome}.
The recent advancements in music source separation~\cite{hennequin2020spleeter, rouard2023hybrid} have enabled singer diarization~\cite{suda2022singer}.
We can directly use speaker diarization models in singer diarization by extracting the vocal tracks from the music.
The previous study~\cite{suda2022singer} adopted a target-speaker voice activity detection (TS-VAD)-based model~\cite{medennikov2020target} trained with a relatively small dataset.
We believe that training the EEND model on a larger data set has the potential to improve singer diarization performance.

Although EEND shows superior performance in speaker diarization, the performance of the model trained without song data deteriorates in singer diarization due to a domain mismatch between speech and song data.
Song data has different formant patterns, speech durations, and singer turn-taking frequencies than speech data.
To address the domain mismatch, a domain adaptation or training from scratch using song data is necessary.
Given few public datasets with singer labels, and the need to cover a variety of musical genres and singers, a simulated data generation method using song data is appealing.
However, most song data contains overlapped choral singing performed by multiple singers or overdubbed as a vocal effect.
If we naively generate training data from the song data containing choral singing data, the choral singing section will be labeled as solo singing.
This mislabeling degrades the performance.

In this paper, we propose a training data generation method based on data cleansing utilizing the neural analysis and synthesis (NANSY++) framework~\cite{choi2022nansy++}.
NANSY++ is a framework to reconstruct an input audio signal based on self-supervised learning and can be used for voice manipulation.
We focus on its strong reconstruction ability derived from pre-training that is specific to the non-overlapped audio signal because of using a large isolated speech dataset.
NANSY++ consists of two processes, analyzing the input audio signal and synthesizing the reconstructed audio signal from the analyzed features.
This characteristic is effective for data cleansing of choral singing.
The input choral singing data is analyzed as a non-overlapped audio signal and the synthesizer generates the solo singing from the analyzed feature. 
In the training data generation for the EEND model, singer labels are given by an energy-based voice activity detection, and then the multiple cleansed song clips are mixed.
The simulated dataset from the cleansed data mitigates the deterioration caused by the mislabeling of automatically generated singer labels.

\begin{figure*}
    \centering
    \includegraphics[width=0.95\linewidth]{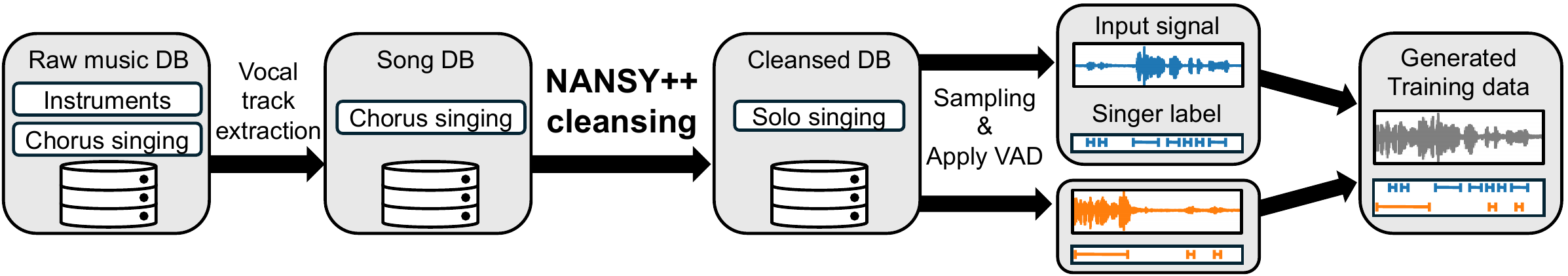}
    \caption{An overview of training data generation process. Our proposed method cleanses choral singing to solo singing. We can generate training data for the singer diarization from the cleansed data.}
    \label{fig:overview}
\end{figure*}

The main contribution of this study is to propose a novel data cleansing method as a new use of the audio signal reconstruction framework based on self-supervised learning.
Our proposed method converts choral singing into solo singing, which mitigates the mislabeling of singer labels for the simulated dataset and enables the training of EEND models for singer diarization.
We conducted an experiment to train the EEND models with internally collected music data and evaluated them with manually annotated 91 popular duet songs.
As a result, our proposed method significantly outperformed baselines, and we confirmed that our proposed method converted choral singing into solo singing.

\section{Background}
\subsection{End-to-End Neural Diarization}
The diarization task is formulated as predicting the speaker/singer labels representing the active duration of each speaker/singer from input audio signal $\mathbf{x}\in\mathbb{R}^{F\times T}$,
where $T$ is the number of frames, and $F$ is the number of feature dimensions.
The EEND model directly predicts sequential speaker label $\mathbf{y} = \{0,1\}^{N \times T}$ as $\hat{\mathbf{y}} = f_\Theta(\mathbf{x})\in [0, 1]^{N \times T}$, where $N$ is the maximum number of speakers and $f_\Theta(\cdot)$ is a function representing the EEND model with parameters $\Theta$.
In the pre-training, the model was trained to minimize binary cross entropy loss $\mathsf{BCE}(y_{t, \phi_n}, \hat{y}_{t, n})$ between ground-truth labels $\{y_{n, t}\}_{n, t=1}^{N, T}$ and predicted labels $\{\hat{y}_{n, t}\}_{n, t=1}^{N, T}$ for each speaker and time frame.
To solve the ambiguity of the permutation of the speakers, permutation invariant training~\cite{kolbaek2017multitalker}, in which only the pair minimizing the loss is calculated as follows:
\begin{align}
    \mathcal{L}(\mathbf{y}, \hat{\mathbf{y}}) = \frac{1}{NT}\min_{\phi \in \mathcal{P}_N}\sum_{n=1}^N\sum_{t=1}^T \mathsf{BCE}(y_{t, \phi_n}, \hat{y}_{t, n}),
\end{align}
where $\mathcal{P}_N$ is the set containing every permutations for $N$ speakers.
This model handles overlapped speech directly and is trained in an end-to-end manner so that the performance can be better than conventional clustering-based diarization models~\cite{garcia2017speaker}.

In singer diarization, the input audio signal is a mixture of singing and instrumental sound.
By extracting a vocal track from the music data using a music source separation model~\cite{hennequin2020spleeter,rouard2023hybrid}, we can apply the EEND models to singer diarization~\cite{suda2022singer}.

\subsection{Neural Analysis and Synthesis Framework}
NANSY++ is a framework that reconstructs the input audio signal based on self-supervised learning~\cite{choi2022nansy++}.
This framework is originally used for voice manipulation e.g., voice conversion or speech/singing voice synthesis by editing intermediate features in the reconstruction.
NANSY++ has two modules, an analyzer and a synthesizer.
The analyzer extracts the analyzed features, e.g., linguistic features, $F_0$, and timbre embedding from an input audio signal, and the synthesizer reconstructs the input audio signal from the analyzed features.
By adopting a large self-supervised model as the backbone network~\cite{baevski2020wav2vec}, the analysis feature does not need to reduce the feature dimension and keeps its powerful representation ability.

The training of NANSY++ is based on self-supervised learning.
In the training, the whole network of NANSY++ is jointly trained with unlabeled audio data.
This enables training with a large amount of isolated speech data and NANSY++ learns strong reconstruction capability.
Considering that NANSY++ does not have mechanisms for overlapped audio signals and training without overlapped audio signals, the pre-trained NANSY++ is specific to non-overlapped audio signals.

\section{Proposed method}
In this section, we propose a data cleansing method utilizing pre-trained NANSY++.
Our proposed method easily generates a large dataset to train EEND models from song data with choral singing.
Fig.~\ref{fig:overview} shows an overview of our proposed method.
Our proposed method cleanses the song data with choral singing and then mixes the cleansed data.

\subsection{Data Cleansing Using NANSY++}
Our proposed method converts choral singing data into solo singing data that can be used to train EEND models.
This conversion is based on the fact that the pre-training of NANSY++ is specific to the non-overlapped audio signal.
The pre-trained analyzer extracts linguistic features, $F_0$, and timbre embedding from the input choral singing as a non-overlapped audio signal.
From these analyzed features, the synthesizer generates solo singing data corresponding to the input song data.
In this cleansing, the sound quality of the singing data degraded, however, that can still be used for training the EEND models.
We use the analyzed feature without any changes for the synthesis because we focus on the data cleansing for the choral singing data.

\subsection{Training Data Generation}
We apply several processes before the cleansing.
At first, we apply music source separation to the original song data as the previous work~\cite{suda2022singer}.
After the vocal track extraction, if the extracted vocal signal has hardly any energy, we remove it as an instrumental track.
Since multiple solo parts by different speakers degrade the quality of our data cleansing because our method is effective only for choral singing, the filtered songs are divided into short segments (30 seconds) to ensure that, at most, one speaker performs a solo part within a segment.

After the cleansing, we mix the cleansed clips.
To generate singer labels, we adopt a simple energy-based voice activity detection (VAD) because it does not require any pre-training for song data.
We split a time-domain cleansed song into $T$-frame chunks $\{s_{l,t}\}^{L,T}_{l,t=1}$, where $L$ is the number of samples in each frame.
The singer labels are obtained by comparing frame-wise energy $e_t$ to threshold $e_\theta$ as follows:
\begin{align}
\label{eq:threshold}
    y_t &=
    \begin{cases}
        1 & \text{if $e_t>e_{\theta}$,} \\
        0 & \text{otherwise},
    \end{cases}\\
    e_t = 10&\log_{10}\left(\frac{\sum_{l}^{L} ||s_{l,t}||^2} 
    {\frac{1}{T}\sum_{t=1}^T\sum_{l=1}^{L} ||s_{l,t}||^2}\right).
\end{align}
After generating the singer label, pairs of the song clips and their corresponding singer label are sampled, and then the input song clips are mixed and the labels are concatenated along the singer
dimension.
In the mixing, the power of the input song clip is randomly sampled, but noise is not added because most music data is recorded in very noise-less conditions.
To increase the variety of mixtures, we adopt dynamic mixing~\cite{zeghidour2021wavesplit}, in which the mixtures are generated randomly on the fly in the training. This method increases the combinations of the song clips and improves the performance compared to generating a finite number of mixtures in advance.

\section{Experimental Evaluation}
We conducted duet singer diarization. In this experiment, we trained EEND models not using manually annotated singer labels for the music data.
We evaluated our proposed method using manually annotated popular duet songs.

\subsection{Dataset}
\label{sec:dataset}
We used multiple datasets to train EEND models and NANSY++.
A summary of the data we used is described in Table~\ref{tab: dataset}.
We generated simulated datasets from the following music and conversational speech data.
We collected $1,027$ hours of internal music clips with singing voice, we call it internal music (IM) data. This dataset contains various singers and genres of music. 
We applied our proposed method to this dataset and generated a simulated dataset to train the EEND models.
We used conversational speech (CS) data commonly used in previous works~\cite{fujita2019end, horiguchi2020end, landini2022simulated}, Switchboard-2 (Phase I, II, III), Switchboard Cellular (Part 1, Part2), and NIST Speaker Recognition Evaluation datasets (2004, 2005, 2006, 2008). 
We generated simulated conversational datasets in the same manner as~\cite{landini2022simulated} from this dataset.
For fair evaluation, we generated two datasets with different overlap ratios.
The overlap ratio of each dataset was $8.1$ and $28.8$, respectively.
The one with the lower ratio was generated using statistics of CALLHOME~\cite{callhome}, which is a telephone conversation data widely used to evaluate speaker diarization performance.
The one with a higher ratio was generated by setting a pause parameter to $0$ to increase the overlap ratio.
This setting led to very frequent overlaps as a conversational simulation.

To train NANSY++, We collected $1,443$ hours of speech data and $12.5$ hours of clean solo singing data.
The amount of the singing voice data was lower than $1\%$ of the amount of the speech data.
The clean speech data includes LibriTTS-R~\cite{koizumi2023libritts}, and an internal Japanese dataset.
The internal dataset contains studio-quality recordings.
The total number of speakers for the speech data was over $4,100$.
In contrast, the number of singers in the solo singing data was only $3$ which was insufficient to generate simulated data. 

We collected 91 popular duet song (PDS) data from CDs. This dataset consisted of 15 male-male songs, 67 male-female songs, and 9 female-female songs.
The average overlap ratio of these data was $62.2$.
Each song contains just two singers who have solo singing parts.
Several songs contained choral singing by an identical singer caused by overdubbing. 
We instruct the manual annotators to label these overlapped singing sections as solo singing.
In addition, we collected VAD labels for sections in which duet singers sing lyrics.
We used these labels in our evaluation because we consider the application of the karaoke systems and we assume these labels are available.

We extracted singing voice from the clips by using Demucs-HT~\cite{rouard2023hybrid}.
The sampling rate was converted to 8 kHz to compare our proposed method to baselines.
The input signals were divided into 30-second clips in the training and not in the evaluation.

\begin{table}
    \centering
    \caption{Summary of the dataset. The third and fourth columns represent the amount of data. }
    \vspace{-3mm}
    \label{tab: dataset}
    \footnotesize
    \begin{tabular}{c|c|cc}
    \toprule
    
    Dataset     & Type  & Song (hour)       & Speech (hour)     \\
    \midrule
    Internal music (IM)              & Train & $1.03\times10^3$  & -                 \\
    Conversational speech (CS)              & Train & -                 & $2.82\times10^3$  \\
    Training data of NANSY++        & Train & $1.25\times10^1$  & $1.44\times10^3$  \\
    \midrule
    Popular duet song (PDS)            & Test  & $6.73\times10^0$  & -                 \\
    \bottomrule
    \end{tabular}
\end{table}

\begin{table*}
    \centering
    \caption{Performance by each method. The arrow in data configurations represents the pre-training dataset and adaptation dataset.
    FA, CF, Under, and Over represent false alarm and confusion, under-counting, and over-counting error respectively. The percentages in the dataset refer to the amount of data.}
    \vspace{-3mm}

    \label{tab:score}
    \begin{tabular}{c|c|r||r|rrr|r|rr}
    \toprule
    \multirow{2}{*}{Method} & \multirow{2}{*}{Dataset} & \multicolumn{1}{c||}{Overlap} & \multicolumn{4}{c|}{DER $\downarrow$} & \multicolumn{3}{c}{D-SCER $\downarrow$} \\
    & & \multicolumn{1}{c||}{ratio (\%)} & Over all & Miss & FA & CF &  \multicolumn{1}{c|}{Over all} & Under & Over \\
    \midrule
    Conv. Sim.~\cite{landini2022simulated}  & CS            
    & $8.1$  & $40.9$ & $29.4$ & $2.3$  & $9.1$  & $22.9$ & $21.2$ & $1.7$ \\
    Conv. Sim.~\cite{landini2022simulated}  & CS            
    & $28.8$ & $38.6$ & $18.7$ & $8.3$  & $11.6$ & $21.9$ & $17.1$ & $4.8$\\
    Naive 2-mix                             & IM ($100\%$)  
    & $59.8$ & $38.2$ & $22.7$       & $2.6$ & $12.8$ & $20.6$ & $18.4$ & $2.2$\\
    \midrule
    Proposed 2-mix                          & IM ($1\%$)    
    & $59.8$ & $24.6$           & $7.3$  & $11.6$ & $5.7$  & $14.1$ & $3.0$ & $11.1$ \\
    Proposed 2-mix                          & IM ($10\%$)   
    & $59.8$ & $23.9$           & $7.1$  & $11.3$ & $5.6$  & $13.9$ & $3.1$ & $10.8$\\
    Proposed 2-mix                          & IM ($100\%$)  
    & $59.8$ & $\mathbf{23.4}$  &  $7.2$ & $10.5$ & $5.7$  & $\mathbf{13.2}$ & $3.2$ & $10.0$\\
    Proposed 2-mix& CS $\rightarrow$ IM ($1\%$) 
    &  $28.8 \rightarrow59.8$  & $24.6$ & $11.8$ & $6.6$  & $6.2$ & $13.5$ & $7.3$ & $6.2$\\
    \bottomrule
    \end{tabular}
    \vspace{-3mm}
\end{table*}

\subsection{Evaluation}
In our experiment, we prepared training data generated by our proposed method and baselines.
As a baseline, we generated a dataset by mixing two song clips without our proposed cleansing, we call it Naive-2mix.
To confirm how the amount of data affects the performance, we generated training data by our proposed method with $10\%$ and $1\%$ subsets of IM data.
In addition, we confirmed the efficacy of the adaptation with our proposed method by additional training of the model trained with CS.

Considering that VAD labels are available in the application of the karaoke system, we focus on evaluations in sections with at least one singer singing.
For this reason, we evaluated our method by the diarization error rate (DER) calculated in only sections with the VAD labels.
This evaluation not only considers the karaoke system but also mitigates the influence of singers except duet singers.

To evaluate singer counting performance independently of the singer identification ability, we introduce duet singer counting error rate (D-SCER) as follows:
\begin{align}
    D\mathchar`-SCER = \frac{Under + Over}{Total},
\end{align}
where the $Under$ and $Over$ are the lengths of sections with under-counting, and over-counting, and $Total$ is the total length of the section where at least one singer is singing in the reference.

\subsection{Conditions}
The model parameters of EEND basically follow the configuration in~\cite{fujita2019end}.
We used the self-attentive EEND model~\cite{fujita2019end} with four-stacked transformer encoders with $256$ dimensional projection layers, $4$ heads, and $2048$ dimensional feed-forward layers.
The input feature was 345-dimensional subsampled log-scaled Mel-filterbank features.
To improve model performance, we adopted intermediate prediction layers and intermediate loss~\cite{yu2022auxiliary}.
We trained each model for $6\times10^{6}$ steps with a batch size of 32.
The optimizer was Adam with the Noam learning rate scheduler and its warmup steps were $2\times10^5$~\cite{vaswani2017attention} for the pre-training.
In the adaptation stage, we fixed the learning rate to $10^{-5}$ and ran $6\times10^{5}$ steps.
The gradient clipping was applied with a norm threshold of $5$.
We applied 11-frame median filtering to prevent production of unreasonably short segments.


We made several changes to NANSY++~\cite{choi2022nansy++} to simplify the implementation while retaining the reconstruction ability.
We replace networks composing the analyzer.
First, we used a reference encoder~\cite{skerry2018towards} to extract a global timbre embedding instead of a time-varying timbre embedding.
The reference encoder extracts $192$-dimensional timbre embedding from $80$-dimensional log-scale mel-spectrograms using six convolutional layers with output channels set to $128$, $128$, $256$,
$256$, $512$, and $512$, respectively.
Second, we replaced the pitch encoder with a signal processing method, and subsequently used the $F_0$ and the voicing/unvoicing flags it produced.
Third, we used a pre-trained ContentVec~\cite{qian2022contentvec} to extract linguistic contents. 
A linguistic encoder converts the $768$-dimensional ContentVec features to $128$-dimensional linguistic features.
In the pre-training, we did not use the information perturbation~\cite{choi2021neural} and the contrastive loss because we focused on the reconstruction ability.
For the synthesizer, we adopted the same model architecture as Period VITS’s decoder~\cite{shirahata2023period}.

In the mixing of our proposed method, clips were mixed keeping the signal-to-noise ratio between $-5$ dB and $+5$ dB.
We set the threshold value in Eq.~\eqref{eq:threshold} to $10$ dB.


\subsection{Results}

Table~\ref{tab:score} shows the performance in the singer diarization.
Our proposed method significantly outperformed the baselines in DER by $14.8$ points and in D-SCER by $8.7$ points.
Compared to Naive 2-mix, our proposed method outperformed and this result shows the effectiveness of NANSY++ data cleansing.
Even if we increased the overlap of CS data, the improvement of DER and D-SCER was limited. 
For this reason, we can say that our method was effective in solving the domain mismatch caused not only by differences in overlap ratio but also by differences in formant patterns.

In terms of the amount of data, the performance of the model trained with $1\%$ was comparable to that of $10\%$ and $100\%$.
A possible reason for the high performance even if the amount of training data was only 10 hours is dynamic mixing that increases the variety of the simulation data. 

The adapted model outperformed the model trained from scratch in over-counting error and deteriorated the under-counting error of D-SCER. 
This result implies that training with our proposed method led to over-counting.

We plot $F_0$ estimated results of raw and cleansed samples of PDS data using~\cite{cuesta2020multiple} and the reference singer labels.
We can see that the higher $F_0$ was removed by our proposed method and the lower $F_0$ was maintained, and we confirmed the choral singing was converted to solo singing by hearing.

\begin{figure}[t]
  \centering
    \includegraphics[width=\linewidth]{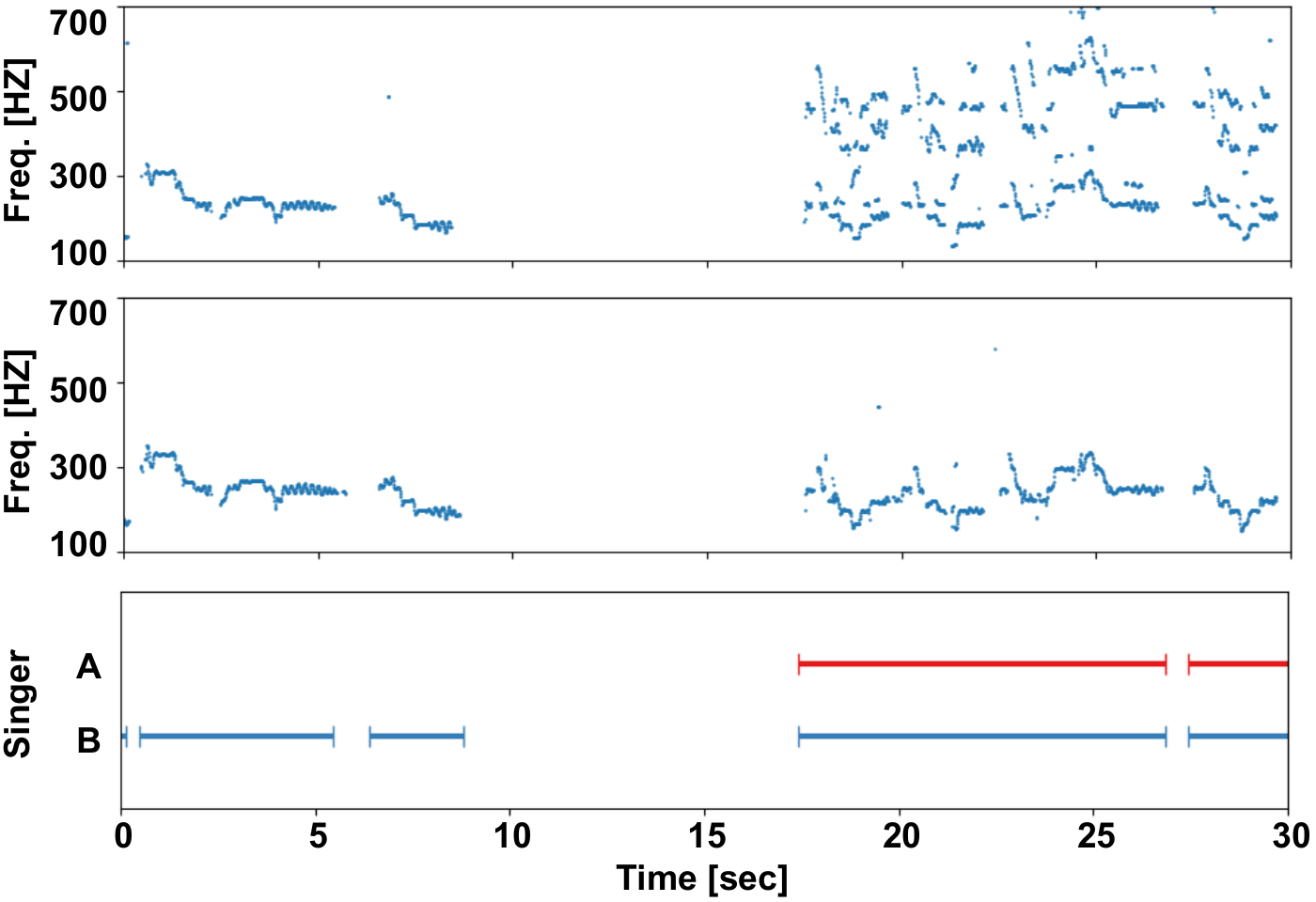}
    \caption{Trajectories of multiple $F_0$ (upper: raw data, lower: cleansed data), and the reference singer labels. The singer A was a female singing with a higher voice and B was a male singing with a lower voice.}
  \label{fig:images}
\end{figure}

\section{Related work}
There are limited prior researches related to the singer diarization~\cite {tsai2004automatic, thlithi2015singer,suda2022singer}.
In previous works, preparing a dataset was challenging.
The latest work~\cite{suda2022singer} trained a model with 53 and evaluated the model with 25 Japanese female idol songs.
Compared to prior works, we evaluated the performance of the model trained with a large dataset which includes various singers and music genres.

To improve the performance under the domain mismatch not using ground-truth labels, semi-supervised adaptation methods are effective~\cite{takashima2021semi, niu2023unsupervised}.
However, the performance of this method depends on an initial model trained in a source domain to train the next models.
For this reason, the performance of these methods is limited under a large domain shift between speaker diarization and singer diarization.

\section{Conclusion}
We proposed a data cleansing method using NANSY++.
Our proposed method can be used to generate a simulated dataset for singer diarization from choral singing data.
Our future work is the further improvement of the diarization performance using semi-supervised adaptation methods. We will apply our proposed method to the initial model of these methods.

\section{Acknowledgement}
We appreciate Kosuke Futamata, Yuma Shirahata, and Hironori Doi collecting data and training NANSY++.

\bibliographystyle{IEEEtran}
\bibliography{mybib}

\end{document}